\documentstyle[prl,aps,epsf,multicol]{revtex}
\begin{document}
\draft

\title{Odd-Frequency Density Waves: Non-Fermi-Liquid
Metals with an Order Parameter}
\author{Eugene Pivovarov}
\address{Department of Physics, Mail Code 103-33, Caltech, Pasadena,
CA 91125}
\author{Chetan Nayak}
\address{Department of Physics and Astronomy, 
University of California, Los Angeles, CA 90095-1547}
\date{\today}
\maketitle

\begin{abstract}
We consider states with a charge-
or spin-density wave order parameter which is odd in frequency,
so that the order parameter vanishes at zero frequency and there is
a conventional Fermi surface. Such states break translational
symmetry and, therefore, are not conventional Fermi liquids. In the
odd-frequency spin-density wave case, there are Goldstone bosons
and the low-energy spectrum is manifestly different from
that of a Fermi liquid. We discuss a simple model which gives rise to such
ordered states. The frequency-dependence of the gap
leads to an unusual temperature dependence for
various thermodynamic and
transport properties, notably the resistivity.
\end{abstract}

\begin{multicols}{2}
\narrowtext

\section{Introduction}

The Fermi liquid is a metallic state which is adiabatically continuable to a
non-interacting electron gas. Its low-energy excitations are in one-to-one
correspondence with those of the non-interacting system. Many attempts to
uncover non-Fermi-liquid metallic behavior in strongly-correlated electron
systems have focused on enhanced scattering mechanisms which might lead to
anomalous behavior in the electron Green function.
In this paper, we attempt a different tack
and focus on metallic states with a conventional Fermi
surface which can be distinguished from Fermi liquids by an {\it order
parameter}. In order to preserve the Fermi surface, the order parameter is
taken to be {\it odd in frequency}\cite{footnote_0}.
When the order parameter breaks a
continuous symmetry, the low-energy
spectrum must also include Goldstone modes.
As a result, odd-frequency ordering is a mechanism by
which the set of low-energy excitations can be expanded from
those of a Fermi liquid (so that they are no longer in one-to-one
correspondence with those of a free Fermi gas), leading to
a manifestly non-Fermi liquid state.

A number of authors have considered superconducting states with order
parameters $\left\langle {c_\alpha}(k,\omega)\, {c_\beta}(-k,-\omega)
\right\rangle$ which are odd in frequency\cite
{Balatsky92,Abrahams93,Abrahams95,Coleman93,Belitz99,Heid95}.
The energetic advantage of such a
state is that it would enable the electrons to
avoid instantaneous Coulomb
repulsion while still benefitting from pairing.
Some exactly soluble
one-dimensional models, such as the one-dimensional Kondo lattice, have a
tendency towards such ordering \cite{Zachar96,Coleman96}.
However, there are claims that the simplest models
of odd-frequency superconducting states suffer
from pathologies which make them unstable \cite{Heid95,Yip}.

In this paper, we consider the analogous states in the particle-hole
channels. We find that simple models of odd-frequency
density wave states do not suffer from any pathologies
(and suspect that this holds for the
corresponding superconducting states, as well).
Odd-frequency states can exhibit a number of interesting
non-Fermi liquid properties including the Goldstone
modes mentioned above,a range of
states above the Fermi surface with finite lifetimes even
in the limit of vanishing frequency and temperature
$\epsilon,T\rightarrow 0$, and a non-mean-field-like
temperature-dependent order parameter.

However, the question of their detection is non-trivial
and cuts to the heart of attempts to experimentally
distinguish non-Fermi liquids from Fermi liquids.
While an odd-frequency superconducting state is, first and foremost,
a superconducting state, which would be identified by its vanishing
resistivity, Meissner effect, etc., an odd-frequency density wave state can
masquerade as a Fermi liquid since the order parameter vanishes at zero
frequency. There will be signatures in thermodynamic and transport
measurements, but they can easily be mistakenly attributed
to other effects, as we discuss below.

\section{Order Parameters and Symmetries}

An odd-frequency charge-density wave state is defined by the anomalous
correlation function 
\begin{equation}  \label{eqn:op-def}
\left\langle {c^{\alpha\dagger}}\left( k,{\epsilon_n}\right) {c_\alpha}
\left( k+Q,{\epsilon_n}\right) \right\rangle = F\left(k,{\epsilon_n}
\right) ,
\end{equation}
where $F\left(k,{\epsilon_n}\right)$ is an odd function of frequency, 
\begin{equation}
F\left(k,-{\epsilon_n}\right) = -F\left(k,{\epsilon_n}\right) .
\end{equation}

To find an equal-time correlation function which serves as an
order-parameter, we Fourier transform (\ref{eqn:op-def}): 
\begin{equation}
\label{eqn:order_time}
\left\langle {T_\tau}\left({c^{\alpha\dagger}}\left( k,\tau\right) {c_\alpha}
\left( k+Q,0\right)\right) \right\rangle = {\tilde F}\left(k,\tau\right) ,
\end{equation}
where ${\tilde F}\left(k,\tau\right)$ is imaginary and odd
in $\tau$. Hence, we can use the
time-derivative as an order parameter: 
\begin{equation} 
\label{eqn:order_time_deriv}
{\left\langle {T_\tau}\left({\partial_\tau}{c^{\alpha\dagger}}\left(
k,\tau\right)\: {c_\alpha}\left( k+Q,0\right)\right)
\right\rangle_{\tau=0}} = 
{\partial_\tau}{\tilde F}\left(k,0\right) .
\end{equation}

The state defined by these order parameters, (\ref{eqn:op-def}) or
(\ref{eqn:order_time_deriv}), breaks translational
symmetry. Time-reversal symmetry is
not broken. This is most easily seen by considering (\ref{eqn:order_time}).
Taking the complex conjugate of both sides of (\ref{eqn:order_time}) gives
\begin{eqnarray}  \label{eqn:order-cc}
\left\langle{T_\tau}\left( {c^{\alpha\dagger}}\left( k+Q,0\right) {c_\alpha}
\left( k,\tau\right)\right) \right\rangle &=& {\left({\tilde F}%
\left(k,\tau\right)\right)^*} \nonumber \\
&=&-{\tilde F}\left(k,\tau\right) .
\end{eqnarray}
Meanwhile, (\ref{eqn:order_time}) is
transformed under time-reversal, ${\cal T}$, into: 
\begin{eqnarray}  \label{eqn:order-t-rev}
\lefteqn {\left\langle{T_\tau}\left({\cal T}\left({c^{\alpha\dagger}}\left(
k,\tau\right) {c_\alpha}\left( k+Q,0\right)\right)\right)\right\rangle }
\nonumber \\
& & \qquad = \left\langle{T_\tau}\left({c^{\alpha\dagger}}\left(
k+Q,0\right) {c_\alpha}\left( k,-\tau\right)\right)\right\rangle \nonumber \\
& & \qquad = -{\tilde F}\left(k,-\tau\right) \nonumber \\
& & \qquad = {\tilde F}\left(k,\tau\right) \nonumber \\
& & \qquad = \left\langle{T_\tau}\left({c^{\alpha\dagger}}\left( k,\tau\right) 
{c_\alpha}\left( k+Q,0\right)\right)\right\rangle ,
\end{eqnarray}
and, hence, the order parameter does not break time-reversal symmetry. In
going from the first equality to the second, we have used
(\ref{eqn:order-cc}).

An odd-frequency spin-density wave is defined by 
\begin{equation}
\left\langle {c^{\alpha\dagger}}\left( k,{\epsilon_n}\right) {c_\beta}\left(
k+Q,{\epsilon_n}\right) \right\rangle = \vec{n}\cdot{\vec{\sigma}%
^\alpha_\beta} F\left(k,{\epsilon_n}\right) ,
\end{equation}
where $F\left(k,{\epsilon_n}\right)$ is again an odd function of frequency
and $\vec{n}$ is the direction chosen spontaneously by the ordered state.
This state breaks translational symmetry and spin-rotational symmetry, which
is broken to the $U(1)$ subgroup of rotations about $\vec{n}$. Again,
time-reversal is preserved. Such a state will exhibit a non-zero
expectations value and anomalous correlations
of the spin nematic order parameter,
${S_i}{S_j}-{\delta_{ij}}{S^2}/3$, such as those
discussed for spin-only models in \cite{Balatsky95}.

\section{Model Interaction}

We now consider a simple two-dimensional model which admits an odd-frequency
charge-density-wave state at the mean-field level. The model contains a
non-singular four-fermion interaction which can be
generated by the exchange of phonons or some gapped electronic collective
mode. For simplicity, we focus on the charge-density-wave case; the
spin-density-wave is analogous.

We consider an effective action which consists of a kinetic term 
\begin{equation}
{S_{\text{0}}} = \int d\tau \int \frac{{d^2}k}{(2\pi)^2}\: {c^{\alpha
\dagger}}(k,\tau) \left({\partial_\tau} - \left(\epsilon(k)-\mu\right)
\right) {c_\alpha}(k,\tau)
\end{equation}
and an interaction term 
\begin{eqnarray}
 \label{eqn:L_int}
{S_{\text{int}}} &=& \frac{1}{\Omega_c^2}\int d\tau {\int_{k,k^{\prime}}}
{\biggl[{c^{\alpha \dagger}}(k+Q){\partial_\tau}{c_\alpha}(k)\biggr]_c} \:
{V_{kk^{\prime}}} \nonumber \\
& & \qquad \times {\biggl[{c^{\beta \dagger}}
(k^{\prime}){\partial_\tau} {c_\beta}(k^{\prime}+Q)\biggr]_c} .
\end{eqnarray}
To avoid clutter, the $\int_{k,k^{\prime}}$ is used as shorthand for the
integrals over $k,k^{\prime}$. The subscript $c$ indicates that the terms in
brackets are actually defined with a frequency cutoff, $\Omega_c$: 
\begin{eqnarray}
\lefteqn {{\biggl[{c^{\alpha \dagger}}(k+Q){\partial_\tau}{c_\alpha}(k)
\biggr]_c}\equiv }\nonumber \\
& & \qquad T {\sum_{|\epsilon_n|<{\Omega_c}}}i{\epsilon_n}\,
{c^{\alpha \dagger}}(k+Q,{\epsilon_n})\,{c_\alpha}(k,{\epsilon_n}) .
\end{eqnarray}
For simplicity, we take ${V_{kk^{\prime}}}$ independent of $k,k^{\prime}$,
${V_{kk^{\prime}}}=\lambda$.
For simplicity, we also take $Q=(\pi,\pi)$ and $\epsilon(k)=
-2t(\cos{k_x}+\cos{k_y})$, corresponding to
commensurate order for a system of electrons on a
square lattice with nearest-neighbor hopping. The generalization to
incommensurate order and other band structures
is straightforward.

\begin{figure}[htb]
\epsfxsize = 1.8 in
\epsffile{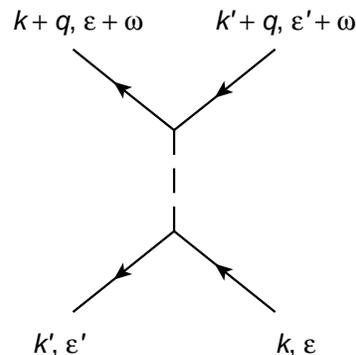}
\caption{A diagram which can lead to an interaction
favorable for odd-frequency density-wave ordering.
The dashed line represents a collective mode
which mediates the interaction.}
\label{fig:int-diag}
\end{figure}

The interaction term ${S_{\text{int}}}$ is long-ranged in precisely the same
way as the BCS reduced interaction. A more realistic short-ranged
interaction would be of the form 
\begin{eqnarray}
{\tilde{S}_{\text{int}}} &=& {\int_{k ,k^{\prime}, q}} {\biggl[{c^{\alpha \dagger}}
(k+q){\partial_\tau}{c_\alpha}(k)\biggr]_c} \:{V^q_{kk^{\prime}}}\nonumber \\
& &\qquad \times {\biggl[{c^{\beta \dagger}}(k^{\prime}){\partial_\tau}
{c_\beta}(k^{\prime}+q)\biggr]_c},
\end{eqnarray}
which includes (\ref{eqn:L_int}) as one term in the sum over $q$. 
Such an interaction could arise from the diagram of figure
\ref{fig:int-diag} if the collective mode has a propagator of the form 
\begin{equation}
\lambda\:\frac{\Omega_c^2}{{\Omega_c^2} +
{v^2}(k-k^{\prime})^2 + ({\epsilon_n}
-{\epsilon_{n^{\prime}}})^2}.
\end{equation}
If $v$ is small, then we can expand the collective mode propagator to obtain 
${V_{kk^{\prime}}}=\lambda$, for $\left| {\epsilon_n}\right| , \left|
{\epsilon_{n^{\prime}}}\right| \leq {\Omega_c}$. Other
terms will also be generated which could drive the formation
of even-frequency order, but they appear to be weaker.

We assume an order parameter of the form: 
\begin{eqnarray}
\alpha \equiv \frac \lambda {\Omega _c^2}\int \frac{{d^2}k}{(2\pi )^2}\,
{\left\langle{c^{\alpha \dagger }}(k+Q)\,i
{\partial _\tau }{c_\alpha }(k)\right\rangle_c}.
\end{eqnarray}
According to our previous observations, $\alpha$
is real.
Then the mean-field action takes the form
\begin{eqnarray}
\label{eqn:action_mf}
{S_{\text{0}}}=T{\sum_n}{\int_k}\,{c^{\alpha \dagger }}(k,{\epsilon _n})
\left( i{\epsilon _n}-\left( \epsilon (k)-\mu \right) \right) {c_\alpha }(k,
{\epsilon _n}) \nonumber \\
-T{\sum_n}\int \frac{{d^2}k}{(2\pi )^2}\:\alpha
{\epsilon _n}\,{c^{\alpha \dagger }}(k,{\epsilon _n}) {c_\alpha }(k+Q,
{\epsilon _n}).
\end{eqnarray}
The equation of motion following from the
mean-field action is:
\begin{equation}
\label{eqn:eom}
\left( i{\epsilon _n}-\epsilon (k)+\mu \right)
{c_\alpha }(k,{\epsilon _n})
- \alpha {\epsilon_n}{c_\alpha }(k+Q,{\epsilon _n}) = 0.
\end{equation}
We multiply the equation of motion by
${c^\dagger_\alpha }(k,{\epsilon _n})$
and take the imaginary time-ordered
expectation value. We see that
the ordinary and anomalous Green functions
satisfy the equation:
\begin{equation}
\label{eqn:FG_eq1}
\left( i\epsilon _n-\epsilon (k)+\mu \right) G(k,\epsilon _n)
-\alpha \epsilon _nF(k,\epsilon _n)=1.
\end{equation}
The right-hand-side results from the time-derivative
(i.e., $i\epsilon_n$) acting on the time-ordering
symbol \cite{comm_rel}.
If we make the replacement $k \rightarrow k+Q$ in
(\ref{eqn:eom}), then we can derive a second equation
in the same way,
\begin{equation}
\label{eqn:FG_eq2}
\left( i\epsilon _n+\epsilon (k)+\mu \right) F(k,\epsilon _n)
-\alpha \epsilon _nG(k,\epsilon _n)=0,
\end{equation}
and the Green function and anomalous Green function for $\left| {\epsilon _n}
\right| \leq \Omega _c$ take the form: 
\begin{mathletters}
\begin{eqnarray}
\label{eqn:FG_exp}
G\left( k,{\epsilon _n}\right)  &=&\frac{i{\epsilon _n}+\mu +\epsilon (k)}
{{\left( i{\epsilon _n}+\mu \right) ^2}-{\left( \epsilon (k)\right) ^2}-
{\alpha ^2}{\epsilon _n^2}}\,, \\
F\left( k,{\epsilon _n}\right)  &=&\frac{\alpha {\epsilon _n}}{{\left( i
{\epsilon _n}+\mu \right) ^2}-{\left( \epsilon (k)\right) ^2}-
{\alpha ^2}{\epsilon _n^2}}\,.
\end{eqnarray}
\end{mathletters}
One might naively think
that equation (\ref{eqn:FG_eq2}) could be obtained by
inspection from (\ref{eqn:FG_eq1}) by replacing
$\Delta({\epsilon_n})$ by ${\Delta^*}(-{\epsilon_n})$
as one usually does in the case of an even-frequency
gap. In this case, this would amount to the
replacement of $\alpha\epsilon_n$
by $-\alpha\epsilon_n$, as was done in Ref.\ \cite{Heid95,Yip}.
However, in the case of an odd-frequency gap,
this simple substitution only works for {\it real}
frequencies. Since the gap is linear in frequency and the
squared modulus of the gap appears in the Green
functions, the Green functions are no longer analytic
in the frequency. As a result,
the analytic continuation from
real to Matsubara frequencies is subtle.
The safe route is to derive both
(\ref{eqn:FG_eq1}) and (\ref{eqn:FG_eq2})
directly from the mean-field action,
as we have done. The naive, incorrect form of
(\ref{eqn:FG_eq2}) would lead to a negative superfluid
density in the case of odd-gap superconductors.
On the other hand, the correct
analogue of (\ref{eqn:FG_eq2}) for a
superconducting action induced by disorder
was used in Ref.\ \cite{Belitz99}.

For $\left| {\epsilon _n}\right| >\Omega _c$, $F$ vanishes and $G$ returns
to its normal state form. In principle, we should also allow for a
quasiparticle renormalization $Z$ resulting from the interaction, but this
does not qualitatively modify our results, so we drop this correction for
simplicity.

{}From these Green functions, we see that, at $\mu =0$, the odd-frequency
charge-density-wave order parameter modifies the quasiparticle spectrum to: 
\begin{equation}
E(k)\equiv \frac{\epsilon (k)}{\sqrt{1+{\alpha ^2}}},
\end{equation}
i.e., it renormalizes the effective mass. For $\mu \neq 0$, the
effect is more complicated. As a result of the odd-frequency
charge-density-wave, $\epsilon
(k)-\mu $ is replaced with 
\begin{equation}
E(k)\equiv \frac{-\mu \pm \sqrt{\left( 1+\alpha ^2\right) {\epsilon ^2}
\left( k\right) -\alpha ^2\mu ^2}}{1+\alpha ^2}. \label{eqn:qp-spectrum}
\end{equation}
{}From (\ref{eqn:qp-spectrum}), we see that the Fermi surface
is unmoved, i.e., $E(k)=0$ when $\epsilon(k)=\mu$,
as we expect, since the order parameter vanishes
at zero frequency.

Furthermore, there is a range of $k$ values above
the Fermi surface, $\left| \epsilon (k)\right|
<\left| \alpha \mu /\sqrt{1+\alpha^2}\right|$,
where $E(k)$ has an imaginary part,
so that quasiparticles in this region have
a finite lifetime. However, these states have zero occupation number,
as the corresponding poles in the Green functions turn out to be lying
outside of the integration contour. This gives rise to a possibility
of the ground state in which some of the quasiparticles are in
levels which are in disconnected from the rest of the Fermi sea.

\section{Gap Equation}

We must now impose a self-consistency condition on $F\left( k,{\epsilon _n}
\right) $, which is the gap equation. We will also impose a
condition on the particle number, thereby implicitly fixing the chemical
potential. These conditions read: 
\begin{mathletters}
\begin{eqnarray}
-\frac \lambda {\Omega _c^2}2T{{\sum_n}^{\prime }}{\int_k}\frac{\epsilon _n^2}
{{\left( i{\epsilon _n}+\mu \right) ^2}-{\left( \epsilon (k)\right) ^2}-
{\alpha ^2}{\epsilon _n^2}} &=& 1,  \label{eqn:gap-eqn} \\
T{{\sum_n}^{\prime }}{\int_k}\frac{i{\epsilon _n}+\mu +\epsilon (k)}{{\left(
i{\epsilon _n}+\mu \right) ^2}-{\left( \epsilon (k)\right) ^2}-{\alpha ^2}
{\epsilon _n^2}} &=& n.  \label{eqn:mu-eqn}
\end{eqnarray}
\end{mathletters}
For simplicity, we consider the case
of half-filling, $n=1$. We have repeated our calculations at
non-zero doping and found similar results.

The prime on the Matsubara summations in (\ref{eqn:gap-eqn}), (\ref
{eqn:mu-eqn}) indicate that they are done with $\alpha=0$
for $\left| {\epsilon_n}\right|> {\Omega_c}$ and
$\alpha\neq 0$ only for $\left| {\epsilon_n}\right|\leq {\Omega_c}$.
A more realistic model replaces the interaction with
one that has a ``smooth'' cutoff ${s_\eta}({\epsilon_n})\, {s_\eta}
({\epsilon_{n^{\prime}}})$ and $\alpha$ by $\alpha\,{s_\eta}({\epsilon_n})$,
with ${s_\eta}(\epsilon)=1$ for $\epsilon\ll {\Omega_c}$ and ${s_\eta}
(\epsilon)=0$ for $\epsilon\gg {\Omega_c}$. We can vary $\eta$ between the
limit $\eta \rightarrow \infty $, which corresponds to a sharp cutoff, and
$\eta \rightarrow 0$ which corresponds to the absence of a cutoff.
For computational simplicity, we take
${s_\eta}(\epsilon)={n_F}(|\epsilon|-{\Omega_c};\beta=\eta)$.

We now discuss the analysis of (\ref{eqn:gap-eqn}), (\ref{eqn:mu-eqn}).
Let us first consider the case of a sharp cutoff.
The left-hand side of the gap equation vanishes if the temperature is above 
\begin{equation}
{T_c^{\text{max}}}=\frac{\Omega_c}\pi .
\end{equation}
${T_c^{\text{max}}}$ is the highest possible transition temperature for this
system. Just below this temperature there is only one pair of terms in the
Matsubara sum which is allowed by the cutoff $\Omega_c$.
As the temperature is decreased, more Matsubara
frequencies begin to contribute, resulting in minor
steps in the phase diagram.

For large $\lambda$, ${T_c} = T_c^{\text{max}}$. Decreasing $\lambda$, we
enter a regime, ${\lambda_{c2}}<\lambda<{\lambda_{c1}}$, in which the system
is in the odd-frequency density-wave phase for an intermediate
range of temperatures ${T_{c2}}<T<{T_{c1}}$.
${\lambda_{c1}}$ is the location of the quantum phase transition at which
the odd-frequency density-wave order
first appears at zero temperature. It is
obtained from (\ref{eqn:gap-eqn}) by setting $\mu =0$ and $\alpha =0$ and
converting the Matsubara sum into an integral: 
\begin{equation}
\int \frac{d^2k}{\left( 2\pi \right) ^2}\left[ \Omega _c-\epsilon \left(
k\right) \arctan \frac{\Omega _c}{\epsilon \left( k\right) }\right] =\frac
{\pi \Omega _c^2}{\lambda _{c1}}.  \label{eqn: lambda_c1}
\end{equation}
The integrand is approximately ${\Omega_c}$ for small
$\epsilon \left( k\right) $ and ${\Omega_c^3}/3{\epsilon^2} \left( k\right)$
for $\left| \epsilon \left( k\right) \right| \gg \Omega _c$.

\begin{figure}[htb]
\epsfxsize =3.375 in
\epsffile{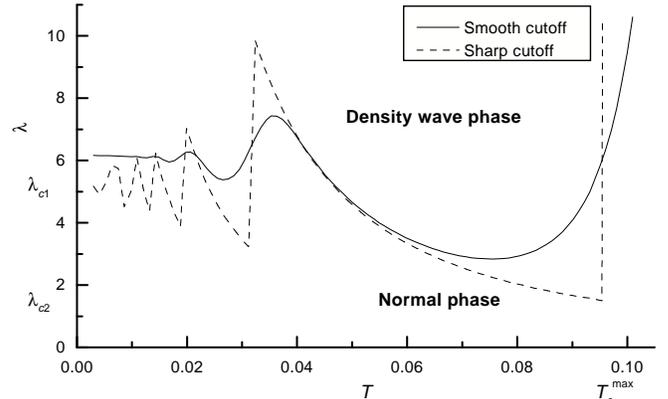}
\caption{Phase diagram at half-filling for ``smooth'' and ``sharp'' cutoffs
at $\Omega =0.3$. The energy scale is set by the width of the band $W=2$.}
\label{fig:odd-phase-diag}
\end{figure}

As $\lambda $ is further decreased, $T_{c2}$ increases and finally reaches
$T_c^{\text{max}}$ at $\lambda _{c2}$. To find $\lambda _{c2}$, we again set
$\mu =0$ and $\alpha =0$, but now we retain only the pair of terms in the
Matsubara sum corresponding to the frequencies $\pm \Omega _c$: 
\begin{equation}
\int \frac{d^2k}{\left( 2\pi \right) ^2}\frac{T_c^{\text{max}}}{\Omega _c^2+
{\epsilon ^2}\left( k\right) }=\frac 1{\lambda _{c2}}. \label{eqn: lambda_c2}
\end{equation}
Comparing expressions (\ref{eqn: lambda_c1}) and (\ref{eqn: lambda_c2}),
we find that the latter is larger by a factor of $3$ for large $\epsilon
\left( k\right) $; hence, $\lambda _{c2}<\lambda _{c1}$.
For $\lambda>\lambda_{c2}$, $\alpha$ jumps
discontinuously at $T_c^{\text{max}}$. Finally, for
$0<\lambda <\lambda _{c2}$, there are no odd-frequency charge-density-wave
solutions. As a result, the phase diagram has the shape shown in Fig.~\ref
{fig:odd-phase-diag} (dashed line) with re-entrant
transitions for ${\lambda
_{c2}}<\lambda <{\lambda _{c1}}$.

Let us now consider how this picture is
modified when we make the cutoff smooth, as it must
be in a physical system. The sharpness of the steps
which separate the re-entrant transitions depends on
the details of the high-frequency cutoff;
they disappear in the limit that
the cutoff is very smooth.
When the cutoff is relatively sharp, more re-entrant transitions are
possible, and there will be several temperature regions in which an
odd-frequency charge-density wave occurs. However, for a smooth cutoff,
there is typically only one such region. The transition at $T_c^{\text{max}}$
is replaced with a smooth curve $T_{c1}(\lambda )$, at which a second-order
transition takes place. The corresponding phase boundary is depicted by the
solid line in Fig.~\ref{fig:odd-phase-diag}.

\begin{figure}[htb]
\epsfxsize =3.375 in
\epsffile{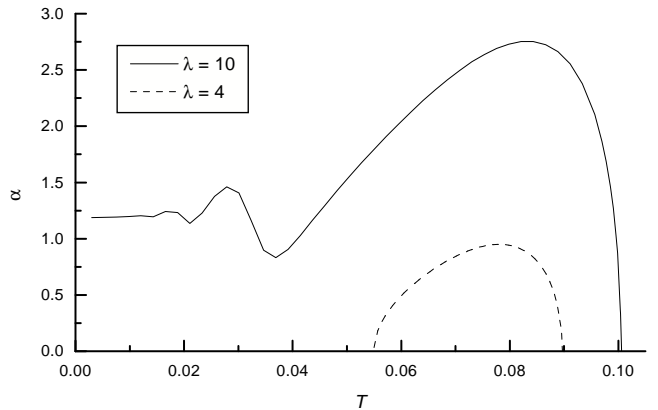}
\caption{Order parameter for different values of the
interaction strength, $\lambda$, with a
``smooth'' cutoff at $\Omega =0.3$.}
\label{fig:alpha-T}
\end{figure}

Below ${T_{c1}}(\lambda )$, $\alpha $ increases as shown in Fig.~\ref
{fig:alpha-T}. Note that for a smooth cutoff all transitions are of second
order, even though the rise of $\alpha $ at $T_{c1}$ is
very steep for large values of $\lambda$ and may give a false impression
of a first-order transition. It is also noteworthy that even
for large $\lambda$ the order parameter $\alpha $ rapidly
attains its maximum as $T$ is decreased below $T_c$
and then decreases as $T\rightarrow 0$ to some non-zero asymptotic value.
This has a significant impact on experimentally measurable
parameters, as we describe in the following section.

The unusual temperature dependence of the order parameter,
which is reflected in the re-entrant phase diagram
and (as we will see in the next section) the condensation
energy, is a consequence of the frequency-dependence
of the gap. In an ordinary, frequency-independent
(or weakly-dependent) ordered state, the condensation energy
at weak-coupling comes primarily from
states near the Fermi energy. At high temperatures,
these states are thermally excited, so there is little
condensation energy to be gained, and the order
parameter decreases as the temperature is increased.
In the case of odd-frequency ordered states,
there is very little condensation energy to be
gained from the particles near the Fermi
surface because their energy is low
(and, hence, they interact weakly with the order parameter).
As a result, the order parameter and condensation energy decrease
as the temperature is decreased.

In fact, there is a second solution to the gap
equation in which it is favorable to have a disconnected
Fermi sea in which some electrons are excited
to a strip in momentum space above
the Fermi surface which is diconnected from
the rest of the Fermi sea, which is centered
about ${\bf k}=0$. When this occurs, the occupation
number does not increase monotonically with
$\epsilon(k)$. However, this solution is higher
in energy, so it does not occur.

\section{Experimental Signatures}

At $T_c$, there will be the usual thermodynamic
signatures of a second order phase transition.
The condensation energy associated with this
transition is the difference between the free energy of the
odd-frequency charge-density-wave state and corresponding free energy
of the normal state at the same temperature. At $\mu=0$ and with a
sharp cutoff, 
\begin{eqnarray}
\label{eqn:cond_energy}
\Delta E\left( T\right) &=& 
\frac{\Omega_c^2}{\lambda}\,{\alpha^2}\:+ \nonumber \\
& & 2T\sum_{n=-n_c-1}^{n_c}{\int_k} \,
\ln\left(\frac{\epsilon (k)^2+ \epsilon _n^2}
{\left( 1 + \alpha ^2\right)
\epsilon _n^2+\epsilon (k)^2}\right) ,
\end{eqnarray}
where $n_c=\left( \Omega _c/\pi T-1\right) /2$.
This equation is obtained by using $\alpha$ as a
Hubbard-Stratonovich field to decouple (\ref{eqn:L_int}),
resulting in the first term in (\ref{eqn:cond_energy}).
The electronic action is then the mean-field action
(\ref{eqn:action_mf}), so that the partition function may be
evaluated to give the second term in (\ref{eqn:cond_energy}).
At zero temperature, the integrals may be evaluated
analytically:
\begin{eqnarray}
\label{eqn:cond_energy_0}
\Delta E\left( 0\right) &=& 
\frac{\Omega_c^2}{\lambda}\,{\alpha^2}\:+
\frac{2}{\pi}{\int_k}{\epsilon(k)}\,
\arctan\left(\frac{\Omega_c}{\epsilon(k)}\right) \nonumber \\
& & -\frac{2}{\pi}{\int_k}\frac{\epsilon(k)}{\sqrt{1+{\alpha^2}}}\,
\arctan\left( \frac{{\Omega_c} \sqrt{1+{\alpha^2}} }{\epsilon(k)}\right)
\nonumber \\
& & +{\Omega_c}\,\frac{2}{\pi}{\int_k}
\ln\left(\frac{{\Omega_c^2}+{\epsilon^2(k)}}
{\left( 1+\alpha ^2\right) {\Omega_c^2}+{\epsilon^2(k)}}\right) .
\end{eqnarray}
The last term is overwhelmingly negative, as may be seen
in various limits (e.g., $\alpha\ll 1$ or ${\Omega_c}\rightarrow\infty$).
Note that the energetic gain comes not from the terms
in the frequency sum (\ref{eqn:cond_energy}) with small Matsubara
frequency, which actually increase energy,
but from the terms near the cutoff -- in a complete
reversal of the situation for a frequency-independent gap\cite{Energy}.

\begin{figure}[htb]
\epsfxsize =3.375 in
\epsffile{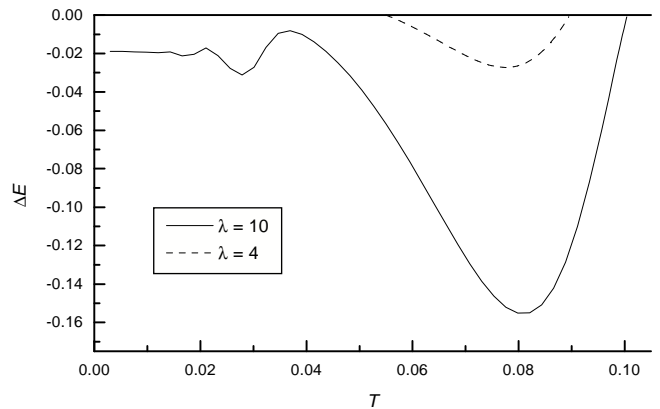}
\caption{Condensation energy as a function of temperature
in model (\ref{eqn:L_int}) with a ``smooth'' cutoff at
$\Omega=0.3$.}
\label{fig:energy-T}
\end{figure}

Again, for a smooth cutoff
the Matsubara frequency sum becomes infinite and $\alpha $ should be
replaced with ${s_\eta }({\epsilon _n})\alpha $. For smooth cutoff
and finite temperature, the condensation energy must be
evaluated numerically.
The dependence of the
condensation energy $\Delta E$ on temperature is shown in Fig.~\ref
{fig:energy-T}. Unlike in even-frequency phases, where $|\Delta E|$ slowly
increases as the system cools down and attains its maximum at zero temperature,
in our model it rapidly reaches a maximum and decreases to a constant
asymptotic value as $T\rightarrow 0$.
As may be seen in figure \ref{fig:energy-T},
the condensation energy is of order of $N_FT_{c1}^2$ at the
maximum, which is comparable to the maximum condensation energy attained
in even-frequency phases.

\begin{figure}[htb]
\epsfxsize =3.375 in
\epsffile{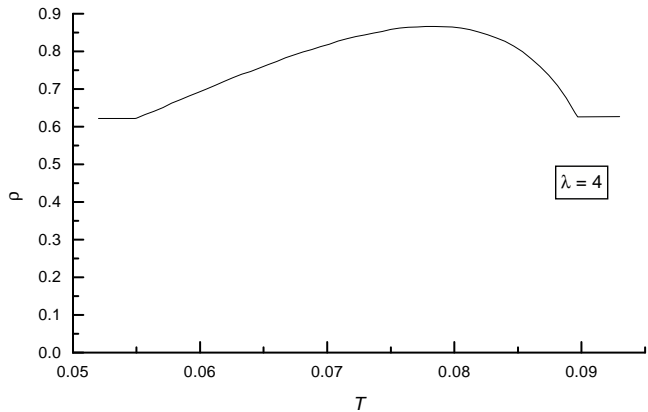}
\caption{Resistivity as a function of temperature
in the model of Eq.\ (\ref{eqn:L_int})
with a ``smooth'' cutoff at $\Omega =0.3$
and interaction strength $\lambda =4$.}
\label{fig:cond}
\end{figure}

The DC conductivity can be computed from the Kubo formula. We assume a
model in which impurities give rise to a lifetime, $\tau $. The resulting
conductivity is given by
\begin{equation}
\sigma =\frac 1{1+\alpha ^2}\int_k\left( \frac{\partial
\epsilon (k)}{\partial k_x}\right) ^2\frac \tau {4T}\frac 1{\cosh \left( 
\frac{E(k)}{2T}\right) ^2},
\end{equation}
where $E(k)$ is given in (\ref{eqn:qp-spectrum}). For small $\mu $, the
modification of the quasiparticle spectrum in the odd-frequency
charge-density-wave phase reduces to the rescaling of the electron
mass near the Fermi surface, so that the new effective mass is
$m^{*}=m\sqrt{1+\alpha ^2}$. This mass enhancement leads to a noticeable
increase of the resistivity,
as shown in Fig.~\ref{fig:cond}. Of course, outside of the the region
${T_{c2}(\lambda )}<T<{T_{c1}(\lambda )}$ the resistivity is that
of the normal phase.

\section{Discussion}

In this work we have considered a model with an odd-frequency
charge-density-wave solution. The
transition to this state is signaled by a second-order phase transition
with a jump in the specific heat. For strong interactions,
the model is in an odd-frequency
charge-density-wave phase for all temperatures $T<T_c$.
For moderately
weak interactions, the model is in such a phase for an intermediate
temperature regime ${T_{c1}}<T<{T_{c2}}$.
There is a quantum phase transition at $\lambda=\lambda_{c1}$;
for weaker interactions, the system is not ordered
at $T=0$.

A similar model admits an odd-frequency spin-density-wave ground state. Such
a state will have, in addition to its Fermi-liquid-like quasiparticles,
Goldstone boson excitations. As a result
${S_i}{S_j}-{\delta_{ij}}{S^2}/3$,
has a non-zero expectation value, and
its correlation functions have Goldstone
poles.

The method that we used to derive the Green functions
of the odd-frequency density wave can be applied
to the analysis of odd-frequency superconductivity as well, with some
modification. The corresponding mean field theory should describe a stable
state with positive superfluid density.

Odd-frequency charge-density wave order results
in mass enhancement.
This affects transport properties; in the density-wave state
the resistivity is considerably larger than in the normal state.
The effect is largest at intermediate temperatures.
At low temperatures, the system is either
in the normal state (for $\lambda < \lambda_{c1}$)
or in an ordered state (for $\lambda \geq \lambda_{c1}$)
with some asymptotic value of the order parameter.
This type of non-monotonic resistivity curve
has been observed in a number of strongly-correlated
electron systems. In layered materials such as the
cuprates and ruthenates, it has been observed
in $c$-axis transport \cite{c-axis}. In
2DEGs, this type of behavior has been observed
in the vicinity of a putative
metal-insulator transition \cite{Experiments}.
It would be premature to suggest that odd-frequency
order is developing in any of these experiments,
but it is noteworthy that it does provide a natural
explanation of otherwise puzzling behavior.

Odd-frequency density wave ordering is also
manifested in thermodynamics. Unlike in even-frequency
states, where the condensation energy $|\Delta E|$ is small near the phase
transition and reaches a maximum at zero temperature, in odd-frequency states
the maximum of $|\Delta E|$ is located near the upper critical temperature.
Consequently, there is strong variation of all thermodynamic quantities with
temperature just below $T_c$. Again, at lower temperatures these phenomena
disappear.

The model that we have introduced is
certainly simplified, as it ignores the
possible proximity of other phases.
However, we believe that odd-frequency density-wave order can
result when a tendency towards ordinary even-frequency density-wave order is
frustrated by some competing interaction. The resolution of the competition
will depend on the scales at which the various interactions act -- essentially
$\Omega_c$ in our model. In particular, one can imagine
a scenario in which an
ordinary even-frequency density wave is
favorable at higher temperatures, but below a
certain temperature, the system undergoes a transition
into an odd-frequency state.

\begin{acknowledgments}
We would like to thank E. Abrahams for discussions.
C.N. is supported by the
National Science Foundation under Grant No.\ DMR-9983544
and by the Alfred P. Sloan Foundation. E.P. is supported
in part by the Department of Energy under grant
DE-FG03-ER-40701.
\end{acknowledgments}


\begin{references}
\bibitem{footnote_0} {A frequency-independent
order parameter may have Fermi pockets, but these are to
be contrasted with the large true Fermi surface
of a Fermi liquid. An odd-frequency
ordered state has a Fermi surface of the latter variety.}
\bibitem{Energy} {The derivation of the
``anomalous'' last term in (\ref{eqn:cond_energy_0})
is straightforward from the partition function
using the path integral formalism.
It is much more difficult, if at all possible,
to explain its presence within
classical thermodynamics.
It is manifestly a quantum phenomenon.}
\end{references}

\end{multicols}

\end{document}